\documentclass[amsmath,amssymb,
onecolumn,apm]{revtex4-2}

\usepackage{graphicx}
\usepackage{caption}
\usepackage{xcolor}
\usepackage[utf8]{inputenc}
\usepackage{rotating}
\usepackage{subcaption}

\DeclareUnicodeCharacter{2212}{-}

\begin{document}

\title{Revisiting the energy-momentum squared gravity}

\author{Mihai Marciu}
\email{mihai.marciu@drd.unibuc.ro}
\affiliation{ 
Faculty of Physics, University of Bucharest, Bucharest-Magurele, Romania
}

\date{\today}

\begin{abstract}
In this paper we have revisited the energy-momentum squared gravity theory, by taking into account the second derivative of the matter Lagrangian with respect to the metric, encapsulating relations originated from thermodynamical grounds. After obtaining the scalar tensor representation of the energy-momentum squared gravity with the new corrections, we have analyzed the physical implications by relying on the linear stability theory. The results show that the current cosmological system is compatible with the expansion of the Universe for some specific matter Lagrangians, explaining the emergence of matter domination era, approaching the late time accelerated expansion era close to the de-Sitter phenomenology.   
\end{abstract}

\maketitle

\section{Introduction}
\label{intro}

\par 
The cosmic picture in the modern cosmology \cite{Nojiri:2017ncd, Bamba:2012cp, Marsh:2015xka} has two fundamental and distinct paradigms. The first paradigm is represented by the dark matter component \cite{Feng:2010gw}, a curious and intriguing phenomena that acts on local and galactic scales and affects the corresponding dynamics \cite{Blumenthal:1984bp, Bertone:2016nfn, Tulin:2017ara, Arias:2012az}. From the observational point of view, there are many studies which analyze the implications of such an unknown component which interacts mainly by the gravitational force \cite{Liddle:1993fq, Lewin:1995rx, Mateo:1998wg, Lesgourgues:2006nd, Graham:2015ouw}. The second paradigm is represented by the dark energy component \cite{Copeland:2006wr, Linder:2002et, Carroll:2003wy}, another fascinating phenomena which acts beyond the galactic scales driving the acceleration of the known universe, affecting the historical evolution \cite{ParticleDataGroup:2018ovx, SDSS:2014iwm, Riess:2006fw}, dominating the cosmic picture in terms of density parameters. Recently, many studies have confirmed this phenomena and analyzed the physical features in different cosmological theories \cite{Planck:2018vyg, WMAP:2003elm, SupernovaSearchTeam:2004lze, Lewis:2002ah}. In these theories, the $\Lambda$CDM model represented the fundamental approach which is compatible with the current evolution of the universe beyond galactic scales \cite{Peebles:2002gy, Nojiri:2010wj, Padmanabhan:2002ji}. Although this model can explain the acceleration of the universe it suffers from various pathological aspects which triggered the appearance of different cosmological theories which extended or complimented the Einstein--Hilbert action in several manners \cite{Capozziello:2011et}. A specific direction in these theories is represented by the modified gravity theories \cite{Nojiri:2006ri} where the Einstein-Hilbert model is replaced by specific actions based on different geometrical invariants \cite{Nojiri:2003ft, Marciu:2021rdl, Marciu:2023hdb, Marciu:2023jvs}. The latter theories can be complimented by scalar fields \cite{Tsujikawa:2013fta, Marciu:2018oks, Vikman:2004dc}, which triggers the accelerated expansion of the universe. These fields can be either canonical of non-canonical, minimally or non-minimally coupled \cite{Marciu:2020vve, Marciu:2019cpb, Marciu:2017sji} with specific topological invariants, affecting different evolutionary aspects \cite{Marciu:2022wzh, Marciu:2022rsc, Marciu:2020yaw}.  
\par 
The most simple modified gravity theory is represented by the $f(R)$ model \cite{DeFelice:2010aj, Nojiri:2017ncd}, a setup that extends the Einstein-Hilbert action by adding a function which depends on the scalar curvature, encoding specific physical features of the manifold \cite{Starobinsky:2007hu}. This theory has been analyzed in various papers \cite{Nojiri:2006gh, Bertolami:2007gv, Cognola:2007zu, Capozziello:2006dj}, further establishing a direction in modified gravity theories towards generic models. In these models, a special class is dedicated to specific theories which are encapsulating various features of the matter component, by encoding an interplay between matter and geometry \cite{Harko:2011kv}. The first generalization is represented by the $f(R, L_m)$ theories \cite{Harko:2010mv}, a natural extension towards a more complete theory of gravity which encompass the matter Lagrangian, encapsulating physical effects due to the specific form of the matter component \cite{Jaybhaye:2022gxq}. To this regard, dynamical features are embedded in these models in view of several physical properties of the matter fluid, described by the Lagrangian and the equation of state \cite{Bhardwaj:2025xff}. A further generalization is represented by the $f(R,T)$ models \cite{AraujoFilho:2025hnf, Mohan:2025uev, Mishra:2025msl, Koussour:2024glo, Pinto:2022tlu}, another class of models which takes into account the trace of the energy-momentum tensor. Furthermore, the $f(R, T^2)$ theory was proposed \cite{Katirci:2013okf,Roshan:2016mbt}, a cosmological framework based on the squared representation of the energy-momentum tensor \cite{Board:2017ign,Akarsu:2018aro}. This theory was studied intensively in the past years, leading to the appearance of various studies \cite{Nazari:2022xhv,HosseiniMansoori:2023zop,Akarsu:2023agp, Sharif:2024fli, Dunsby:2025ahd, Pereira:2024kmj, Marciu:2023muv, Moraes:2017dbs, Nari:2018aqs, Sharif:2025aul, Siddiqi:2025vwv, Sharif:2024ism, Fu:2024cjj}. The furthest class of modified gravity theories which treat matter and geometry on equal footing is described by the $f(R, T_{\mu\nu}G^{\mu\nu})$ cosmologies \cite{Marciu:2024gqv, Ayuso:2014jda, Cipriano:2023yhv, Haghani:2013oma, Odintsov:2013iba, Shahidi:2025eng, Sharif:2013kga} based on the interplay between specific features of the geometrical manifold and physical attributes of the matter component, encoded into the representation of the energy-momentum tensor.  

\par 
In the derivation of the gravitational field equations for the energy-momentum squared gravity theory \cite{Cipriano:2024jng} one encounters a term that contains the second variation of the matter Lagrangian with respect to the metric. In the general case \cite{Bahamonde:2019urw} this term has been neglected, leading to an approximate theory. Recently, a new paper \cite{Akarsu:2023lre} has analyzed the second variation of the matter Lagrangian with respect to the metric, showing that the inclusion of such a term can lead to several physical effects. In the present paper, we shall extend the analysis of the energy-momentum squared gravity \cite{Bahamonde:2019urw} by taking into consideration into the analysis the second variation of the matter Lagrangian with respect to the metric. 
\par 
The plan for present paper is the following. In Secs.~\ref{background}, \ref{actiune} we review the energy-momentum squared gravity theory, discussing the emergence of the gravitational field equations. The focus is on the inclusion of the second variation of the matter Lagrangian with respect to the metric. Then, in Sec.~\ref{actiune} we obtain the scalar tensor representation of the latter theory for two specific cases, depending of the form of the matter Lagrangian. Furthermore, in Secs.~\ref{ds1}--\ref{ds2} we apply the dynamical analysis and analyze the phase space structure, discussing the physical aspects of the cosmological system. In Sec.~\ref{nonex} we discuss some asymptotical solutions for the previous cosmological models. Lastly, in Sec.~\ref{conclusions} we make a short summary and give the final concluding remarks.

\section{Fundamental relations for the energy-momentum-squared gravity}
\label{background}
\par

Before proceeding to the analysis of the dynamical properties of the energy-momentum-squared gravity in the scalar tensor representation we shall first introduce some key concepts and features related to the second derivative of the matter Lagrangian with respect to the metric. The exposition is based on a recent work by Ozgur et al. \cite{Akarsu:2023lre} in the context of modified gravity theories. The analysis starts with the assumption that the matter component of the Universe (as the baryonic matter and the dark matter system) can be represented by a perfect fluid having the following energy momentum tensor:

\begin{equation}
    T_{\mu\nu}=(\rho+p)u_{\mu}u_{\nu}+p g_{\mu\nu},
\end{equation}

with $\rho$ the energy density, $p$ the pressure, and $u_{\mu}$ the four-velocity of the matter fluid ($u_{\mu} u^{\mu}=-1$). Taking into account that the fundamental theory associated to the matter component is described by a specific Lagrangian obeying the principle of least action, we define the energy-momentum tensor in the usual manner, 

\begin{equation}
    T_{\mu\nu}=-\frac{2}{\sqrt{-g}}\frac{\delta(\sqrt{-g}L_m)}{\delta g^{\mu\nu}},
\end{equation}

with $L_m$ the Lagrangian density of the matter component. If we consider that the matter Lagrangian directly depends on the metric component we can write the following relation, 

\begin{equation}
    T_{\mu\nu}=L_m g_{\mu\nu}-2\frac{\partial L_m}{\partial g^{\mu\nu}}.
\end{equation}

In the modified gravity theories the matter Lagrangian for the perfect fluids is not uniquely defined. Hence, we have the following cases which shall be considered. In the first case the matter Lagrangian is defined by the corresponding pressure ($L_m=p$), while in the second one is described by the matter density, $L_m=-\rho$. Taking into account the previous aspects, we can write the first derivative of the matter Lagrangian \cite{Akarsu:2023lre}, 

\begin{equation}
    \frac{\delta(L_m=p)}{\delta g^{\mu\nu}}=-\frac{1}{2}(\rho+p)u_{\mu}u_{\nu},
\end{equation}

\begin{equation}
    \frac{\delta(L_m=-\rho)}{\delta g^{\mu\nu}}=-\frac{1}{2}(\rho+p)(u_{\mu}u_{\nu}+g_{\mu\nu}).
\end{equation}

\par
In Ref.~\cite{Akarsu:2023lre} the authors have shown that the above relations obtained for the first derivative of the matter Lagrangian with respect to the metric tensor are compatible with different thermodynamical aspects. 
\par 
In the case where the matter Lagrangian is defined by the pressure ($L_m=p$), the exposition starts with the assumption that the pressure of the matter component is described by the specific enthalpy ($h$), and the corresponding entropy ($s$) per unit mass, $p=p(h,s)$. Starting from the definition of the matter Lagrangian, it is shown that the variation of the action for the matter component is compatible with the first law of thermodynamics, giving the usual definition of the energy momentum tensor in terms of thermodynamical variables. The second derivative of the matter Lagrangian with respect to the metric tensor is described by the following relation \cite{Akarsu:2023lre},

\begin{equation}
    \frac{\partial^2 (L_m=p)}{\partial g^{\mu\nu}\partial g^{\sigma \epsilon}}=-\frac{1}{4}(1-\frac{\partial \rho}{\partial p})(\rho+p)u_{\mu} u_{\nu} u_{\sigma} u_{\epsilon}.
\end{equation}
This relation is viable if one takes into account the conservation of the specific entropy, $\delta s=0$.
\par 
In the second case when the matter Lagrangian is described by the specific density $(L_m=-\rho)$, the exposition starts with the assumption that the density of the perfect fluid depends on the particle number density $(n)$ and the specific entropy $(s)$, $\rho=\rho(n,s)$. Taking into account the conservation of the specific entropy  and the first law of thermodynamics, the variation procedure leads to the usual definition of the matter energy momentum tensor for a perfect fluid in this case. The second derivative of the matter Lagrangian with respect to the metric tensor is deduced in a similar manner \cite{Akarsu:2023lre}, 

\begin{equation}
     \frac{\partial^2 (L_m=-\rho)}{\partial g^{\mu\nu}\partial g^{\sigma \epsilon}}=-\frac{1}{4}(\rho+p)\Big[ (1+\frac{\partial p}{\partial \rho})(u_{\mu}u_{\nu}+g_{\mu\nu})(u_{\sigma}u_{\epsilon}+g_{\sigma \epsilon})+2 (u_{\mu}u_{\nu}u_{\sigma}u_{\epsilon}-g_{\sigma \mu} g_{\epsilon \nu}) \Big],
\end{equation}

valid if the specific entropy is conserved, together with the particle number flux vector density.
\par 
In the dust scenario where the dark matter pressure is zero the second derivative of the matter Lagrangian with respect to the metric tensor contains a problematic term in the first case where the matter Lagrangian is defined by the  pressure ($L_m=p$), leading to a divergence in the corresponding equations. Due to this reason, this term was omitted in the original analysis specific for the dust cosmology. 
\par 
As can be seen in the second case when the matter Lagrangian is described by the specific density $(L_m=-\rho)$, the theory is free from any problematic terms. The inclusion of the second derivative of the matter Lagrangian in the analysis might lead to a more complete theory having different physical effects.

\section{The model and the gravitational field equations}
\label{actiune}
\par 
In this paper we are interesting in revisiting the energy-momentum-squared gravitational theory, by encapsulating physical effects due to the appearance of the second derivative of the matter Lagrangian with respect to the metric. Hence, we consider the following action \cite{Cipriano:2023yhv}, 

\begin{equation}
    S=\frac{1}{2}\int \sqrt{-g} f(R, T^2) d^4 x + \int \sqrt{-g} L_m d^4 x,
\end{equation}

where $T^2=T_{\mu\nu}T^{\mu\nu}$ represents the contraction of the matter energy-momentum tensor, whereas $L_m$ denotes the Lagrangian for the matter component. Notice that in this model the baryonic matter is encapsulated into the dark matter component, described by the definition of the energy-momentum tensor, 

\begin{equation}
    T_{\mu\nu}=-\frac{2}{\sqrt{-g}}\frac{\delta \sqrt{-g} L_m}{\delta g^{\mu\nu}}.
\end{equation}

We shall assume that the matter component is described by a barotropic equation of state, $p_m=w_m \rho_m$. For simplicity, in our computations we shall neglect the subscript $m$. The gravitational equations which governs the dynamics of the current cosmological model are obtained by taking the variation of the total action with respect to the metric (or inverse metric $g^{\mu\nu}$), applying the variational principle. Hence, we arrive at the following equation \cite{Cipriano:2023yhv}, 

\begin{equation}
  f_R R_{\mu\nu}-\frac{1}{2} g_{\mu\nu}f-(\nabla_{\mu}\nabla_{\mu}-g_{\mu\nu}\Box)f_R=T_{\mu\nu}-f_{T^2}\Theta_{\mu\nu},  
\end{equation}

with $\Theta_{\mu\nu}=\frac{\delta (T_{\alpha\beta}T^{\alpha\beta})}{\delta g^{\mu\nu}}$. The conservation equation is deduced by applying the covariant derivative to the field equations, leading to \cite{Cipriano:2023yhv}

\begin{equation}
    \nabla_{\mu} T^{\mu\nu}=\nabla_{\mu}(f_{T^2}\Theta^{\mu\nu})+f_R\nabla_{\mu}R^{\mu\nu}-\frac{1}{2}g^{\mu\nu}\nabla_{\mu}f,
\end{equation}
which leads to the modified continuity equation in the case of the specific cosmological model for the geometrical background, embedded into the form of the corresponding metric. Since the standard continuity equation is not satisfied ($\nabla_{\mu}T^{\mu\nu} \neq 0$), a matter creation or annihilation process can manifest, leading to an interplay between matter and geometry by particle production or absorption. The physical effects which appear due to the second order derivative of the matter Lagrangian are encapsulated into the definition of $\Theta_{\mu\nu}$, valid once the fluid component is introduced. 
    
\par 
In order to study the physical effects in the energy-momentum-squared gravity theory we shall introduce the scalar tensor representation of the latter theory, working in a similar description. In order to obtain the scalar tensor representation we shall introduce two additional scalar fields in the action \cite{Cipriano:2023yhv}, 

\begin{equation}
    S=\frac{1}{2} \int \sqrt{-g} d^4 x (f(\alpha, \beta) +f_{\alpha}(R-\alpha) +f_{\beta}(T^2-\beta)),
\end{equation}
with $f_{\alpha}=\frac{\partial f}{\partial \alpha}$, $f_{\beta}=\frac{\partial f}{\partial \beta}$. In this representation, we notice the existence of three corresponding fields, the gravitational field represented by the metric $g_{\mu\nu}$, and the additional scalar fields $\alpha$ and $\beta$. The variational principle implies the computation of three variations of the total action for the three fundamental fields. For the $\alpha$ and $\beta$ fields we obtain the following equations, 

\begin{equation}
    f_{\alpha \alpha} (R- \alpha) +f_{\beta \alpha}(T^2-\beta)=0,
\end{equation}

\begin{equation}
    f_{\alpha \beta} (R- \alpha) +f_{\beta \beta}(T^2-\beta)=0,
\end{equation}

 a system which has a unique solution if $f_{\alpha \alpha} f_{\beta \beta} \neq f_{\alpha \beta}^2$. This solution implies $\alpha=R$, $\beta=T^2$. Further, the solution is described by the two scalar fields:

 \begin{equation}
     \phi=\frac{\partial f}{\partial \alpha},
 \end{equation}

 \begin{equation}
     \psi=\frac{\partial f}{\partial \beta},
 \end{equation}

 connected by an interaction potential,

 \begin{equation}
     V(\phi, \psi)=-f(\alpha, \beta)+ \phi \alpha +\psi \beta.
 \end{equation}

 The final expression for the scalar tensor representation in the case of energy-momentum-squared gravity is, 

 \begin{equation}
     S=S_m+\frac{1}{2} \int \sqrt{-g}~d^4 x(\phi R + \psi T^2 -V(\phi, \psi)),
 \end{equation} 
 with $S_m$ the action for the matter (dark matter) component. For this action we obtain the following field equations \cite{Cipriano:2023yhv}, 

 \begin{equation}
     \phi G_{\mu\nu}+\frac{1}{2}g_{\mu\nu}V-(\nabla_{\nu} \nabla_{\mu}-g_{\mu\nu} \Box)\phi=T_{\mu\nu}-\psi (\Theta_{\mu\nu}-\frac{1}{2}g_{\mu\nu}T^2),
 \end{equation}

 \begin{equation}
     \frac{\partial V(\phi, \psi)}{\partial \phi}=R,
 \end{equation}

 \begin{equation}
     \frac{\partial V(\phi, \psi)}{\partial \psi}=T^2,
 \end{equation}

 while the conservation equation reduces to:

 \begin{equation}
     \nabla_{\mu}T^{\mu\nu}=\nabla_{\mu}(\psi \Theta^{\mu\nu})-\frac{1}{2}g^{\mu\nu}(R \nabla_{\mu}\phi+\nabla_{\mu}(\psi T^2-V(\phi, \psi)).
 \end{equation}

\par
Now we are in position to obtain the final gravitational field equations in the scalar tensor representation for the energy-momentum-squared gravity, incorporating viable effects from the second order derivative of the matter Lagrangian. For the matter Lagrangian we shall discuss two separate cases, $L_m=p$, and $L_m=-\rho$, respectively. We shall consider the Roberson-Walker metric, $ds^2=-dt^2+a^2(t) \delta_{ij}dx^idx^j$, with $a(t)$ the cosmic scale factor. 
\par 
In the case where $L_m=p$, we obtain the following acceleration equation:

\begin{equation}
    p+\frac{1}{2}\psi(\rho^2+3 p^2)=-\phi(2 \dot{H}+3 H^2)+\frac{1}{2}V+H\dot{\phi}-\ddot{\phi}-3 H \dot{\phi},
\end{equation}

while the Friedmann constraint equation reduces to:

\begin{equation}
    3 H^2 \phi = \rho+\frac{1}{2}V-3 H \dot{\phi}-\psi \rho (\rho+p)(1-\frac{\partial \rho}{\partial p})+\frac{1}{2} \psi (\rho^2+8 \rho p+3 p^2),
\end{equation}

having a divergence in the case where the matter component is described by a dust fluid scenario.

\par 
On the other hand, in the case where $L_m=-\rho$ we have:
\begin{equation}
    p-\psi \rho^2(1+w+3w^2+3w^3)+\frac{1}{2}\psi(\rho^2+3 p^2)=-\phi(2 \dot{H}+3 H^2)+\frac{1}{2}V+H\dot{\phi}-\ddot{\phi}-3 H \dot{\phi},
\end{equation}

\begin{equation}
    3 H^2 \phi= \rho+\frac{1}{2}V-3 H \dot{\phi}-\frac{1}{2}\psi(\rho^2+3 p^2),
\end{equation}

where we have used the definition of the barotropic equation of state $p=w \rho$, ($w=const.$) at constant entropy. Notice that in the latter case for the matter Lagrangian the equations are divergent free in the dust scenario, representing viable relations which incorporate specific thermodynamic assumptions. We mention here that in the literature the matter Lagrangian can have also other forms, depending on the specific considerations \cite{Minazzoli:2025gyw}. Furthermore, the matter equation of state can go beyond the barotropic scenario \cite{Bhat:2024mqz}. 

\section{The dynamical analysis for $L_m=p$}
\label{ds1}
\par 
In this section we shall discuss the physical effects which appear in the scalar tensor representation of the energy-momentum-squared gravity by adopting the linear stability theory. In order to study the physical properties of our model, we first introduce a set of independent variables compatible to our cosmological system:

\begin{equation}
    x=\phi,
\end{equation}

\begin{equation}
    y=\frac{V(\phi, \psi)}{6 H^2},
\end{equation}

\begin{equation}
    z=\frac{\dot{\phi}}{H},
\end{equation}

\begin{equation}
    v=\rho ~ \psi, 
\end{equation}

\begin{equation}
    \Omega_m=\frac{\rho}{3 H^2}.
\end{equation}

\par 
This set of independent variables is complimented by another set of auxiliary variables which are dependent, used in order to obtain the final form of the autonomous system in the first order, 

\begin{equation}
    \Gamma=\frac{\ddot{\phi}}{H^2},
\end{equation}

\begin{equation}
    \Pi=\frac{\dot{\rho}}{H^3},
\end{equation}

\begin{equation}
    \Sigma=\frac{\dot{H}}{H^2},
\end{equation}

\begin{equation}
    E=\dot{\psi}H.
\end{equation}

For the potential we shall adopt a power law representation, $V(\phi, \psi)=V_0 \phi^n \psi^m$, with $V_0, n, m$ constant parameters. In this representation the Friedmann constraint equation reduces to:

\begin{equation}
    x=\frac{v \left(3 w^3+6 w^2+w+2\right) \Omega _m}{2 w}+\Omega _m+y-z,
\end{equation}

with the following acceleration equation:

\begin{equation}
    2 \Gamma +9 v w^2 \Omega _m+3 v \Omega _m+6 w \Omega _m+(4 \Sigma +6) x-6 y+4 z=0.
\end{equation}

For the first equation associated to the potential we obtain:

\begin{equation}
    \Sigma =\frac{n y-2 x}{x},
\end{equation}

while the second one reduces to:

\begin{equation}
    v=\frac{2 m y}{\left(3 w^2+1\right) \Omega _m}.
\end{equation}

Another equation is obtained by differentiating the second equation of the potential, obtaining:

\begin{equation}
    \frac{3 m n y z}{x}+\frac{9 E (m-1) m y \Omega _m}{v}-\Pi  v \left(3 w^2+1\right)=0.
\end{equation}

Lastly, we get another constraint by differentiating the first Friedmann equation, 

\begin{multline}
    -6 \Gamma +9 E \Omega _m^2+\frac{18 E m y \Omega _m}{v}+27 E w^2 \Omega _m^2+54 E w \Omega _m^2
    \\
    +\frac{18 E \Omega _m^2}{w}+z \left(\frac{6 n y}{x}-6 (\Sigma +1)\right)+2 \Pi +2 \Pi  v+6 \Pi  v w^2+12 \Pi  v w+\frac{4 \Pi  v}{w}-12 \Sigma  x=0.
\end{multline}
\par 
Taking into account the previous set of equations, we obtain for the non-independent auxiliary variables the following relations:

\begin{equation}
    \Gamma =\frac{1}{2} \left(-3 \Omega _m \left(3 v w^2+v+2 w\right)+(6-4 n) y-4 z\right)+x,
\end{equation}

\begin{equation}
    \Pi =\frac{3 m y \left(n z \left(v \left(3 w^3+6 w^2+w+2\right) \Omega _m+2 m w y\right)-3 (m-1) w x \left(\Omega _m \left(3 v w^2+v+2 w\right)-2 y+2 z\right)-6 (m-1) w x^2\right)}{x \left(v^2 (w+2) \left(3 w^2+1\right)^2 \Omega _m+2 m y \left(v \left(3 w^2+1\right) (m (w+2)-2)+(m-1) w\right)\right)},
\end{equation}

\begin{equation}
    \Sigma =\frac{n y}{x}-2,
\end{equation}

\begin{equation}
    E=-\frac{v \left(2 v \left(3 w^2+1\right) \left(m n (w+2) y z+3 w x \left(w \Omega _m-y+z\right)+3 w x^2\right)+2 m n w y z+3 v^2 w \left(3 w^2+1\right)^2 x \Omega _m\right)}{3 x \Omega _m \left(v^2 (w+2) \left(3 w^2+1\right)^2 \Omega _m+2 m y \left(v \left(3 w^2+1\right) (m (w+2)-2)+(m-1) w\right)\right)}.
\end{equation}

Considering the translation to the e-folding number $N=log(a)$ in the representation of the autonomous system associated to the independent variables ($y, z, \Omega_m$), we have:

\begin{equation}
    \label{unu1}
    \frac{dy}{dN}=\frac{3}{2} E \left(3 w^2+1\right) \Omega _m^2-2 \Sigma  y+(\Sigma +2) z, 
\end{equation}

\begin{equation}
    \frac{dz}{dN}=\Gamma -\Sigma  z, 
\end{equation}

\begin{equation}
\label{ddoi1}
    \frac{d \Omega_m}{dN}=\frac{\Pi }{3}-2 \Sigma  \Omega _m.
\end{equation}

For this system we have identified the following critical points by setting the r.h.s. of the previous eqs. \eqref{unu1}--\eqref{ddoi1} to zero in a situation where the matter component acts closely as a dust ($w \to 0, w \neq 0$).
\par
\begin{figure}[h]
  \includegraphics[width=8cm]{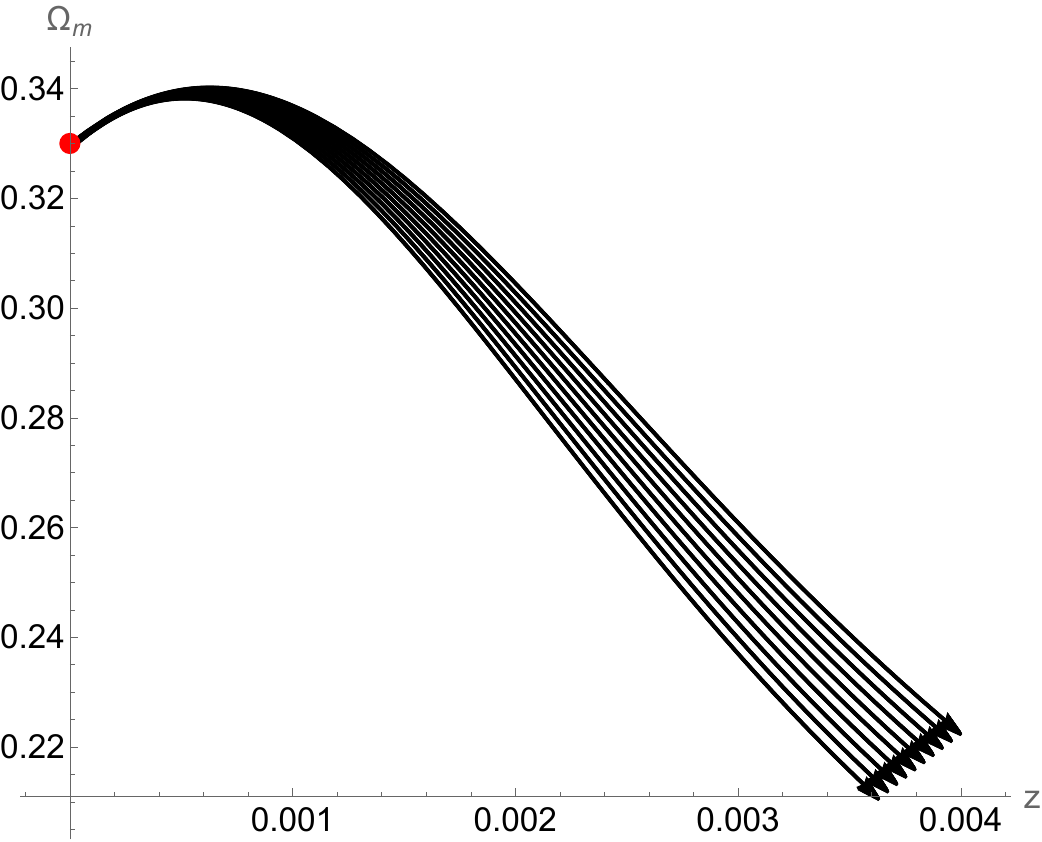} 
\caption{The phase space directions near the $P_1$ solution in the $(z, \Omega_m)$ plane.}
\label{fig:p1fig1}       
\end{figure}

The first solution is represented by a critical line located at the coordinates (found in the case where $m=-1$):

\begin{equation}
    P_1=\Big[y, z=0, \Omega_m=-\frac{2 (n w-3 w+1) y}{w (3 w-1)} \Big],
\end{equation}
with a scaling behavior since $w_{eff}=-1-\frac{2}{3}\frac{\dot{H}}{H^2}=w$. The existence conditions imply that $n w y+n y\neq 0$. The corresponding eigenvalues are the following:

\begin{equation}
  \Big[ 0,\frac{3 (w-1)}{4}\pm \frac{\sqrt{n^2 w^3 \left(6 (4 n-15) w^2+(57-8 n) w+9 w^3-8\right)}}{4 n w^2}\Big],  
\end{equation}
describing a non-hyperbolic solution with one zero eigenvalue. By fine-tuning, the critical behavior can correspond to a saddle solution. For example, if the following conditions are met, 

\begin{equation}
    \left(w<0\land n>0\land \left(n\leq 3\lor \frac{1}{3-n}<w\right)\right)\lor \left(3 w>1\land \left(n\geq 3\lor \left(\frac{1}{3-n}>w\land n>0\right)\right)\right),
\end{equation}
then one eigenvalue has a positive real part, while the other has a negative real part, corresponding to a saddle critical line. Notice that at this solution the kinetic energy of the $\phi$ field is zero, describing a static configuration for the $\phi$ field. The numerical evolution in the phase space structure near the $P_1$ solution is displayed in Fig.~\ref{fig:p1fig1}.

\par
\begin{figure}[h]
  \includegraphics[width=8cm]{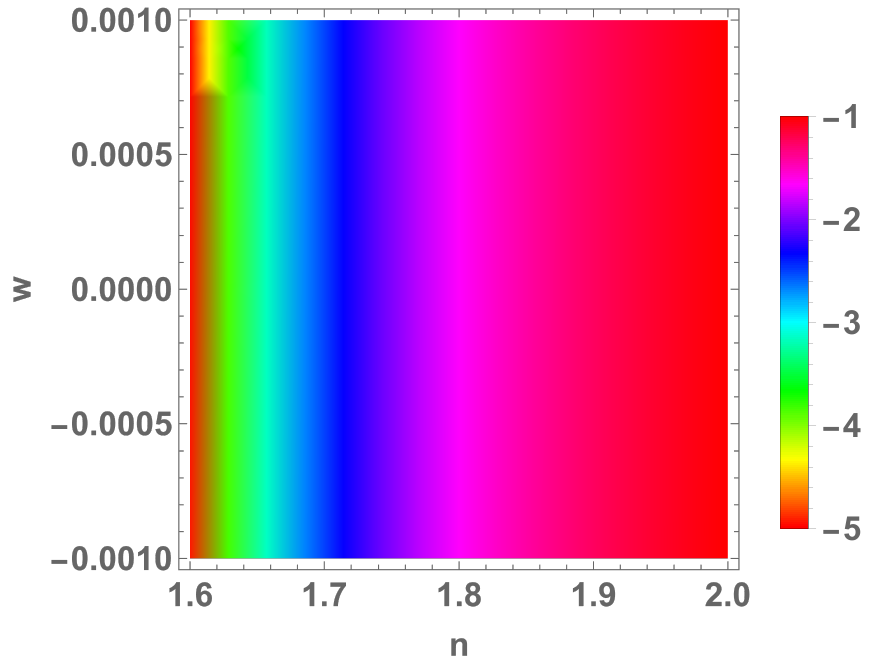} 
\caption{The effective equation of state for the $P_2$ solution. }
\label{fig:p2fig11}       
\end{figure}

 
The second solution found exists in the case where we set $m=\frac{2 w-n w}{w-1}$, with the following existence conditions: $n^2 w^2 y+2 n^2 w y-5 n w^2 y-7 n w y+6 w^2 y+6 w y\neq 0$. In this case the solution is found at the coordinates:

\begin{equation}
    P_2=\Big[y, z=0, \Omega_m=0 \Big],
\end{equation}
another critical line where the potential energy of the two fields is not set to a specific value. In this case the geometrical dark energy component completely dominates in terms of density parameters, with a static configuration for the $\phi$ field. The effective equation of state, 
\begin{equation}
    w_{eff}=\frac{(n-1) w-1}{n (w+2)-3 (w+1)},
\end{equation}
can correspond to an accelerated expansion. For example, if the following non-exclusive conditions are met,
\begin{multline}
    \left(2 n>3\land n\leq 2\land \left(\frac{3}{n-3}+w+2<0\lor w>\frac{3-n}{2 n-3}\right)\right)\lor\left(n>2\land n<3\land \left(w<\frac{3-n}{2 n-3}\lor \frac{3}{n-3}+w+2>0\right)\right),
\end{multline}
then the solution is associated to an accelerated expansion era. A non-exhaustive interval for the effective equation of state in the case of $P_2$ solution is displayed in Fig.~\ref{fig:p2fig11}. The resulting eigenvalues are the following:

\begin{equation}
    \Big[0,\frac{-3 n+3 w+3}{n (w+2)-3 (w+1)},\frac{3 (w+1) ((n-3) w+1)}{w (n (w+2)-3 (w+1))} \Big].
\end{equation}
As in the previous case, the saddle behavior can manifest if the following non-exclusive conditions are considered, 

\begin{multline}
    \left(1<n\leq \frac{3}{2}\land w>0\land n>w+1\right)\lor
    \\ 
    \left(2 n>3\land n<2\land \frac{3-2 n}{n-3}<w\land w+1<n\right)\lor (2<n<3\land n w+1<3 w\land n (w+2)>3 (w+1)).
\end{multline}

\par 

\begin{figure}[h]
  \includegraphics[width=8cm]{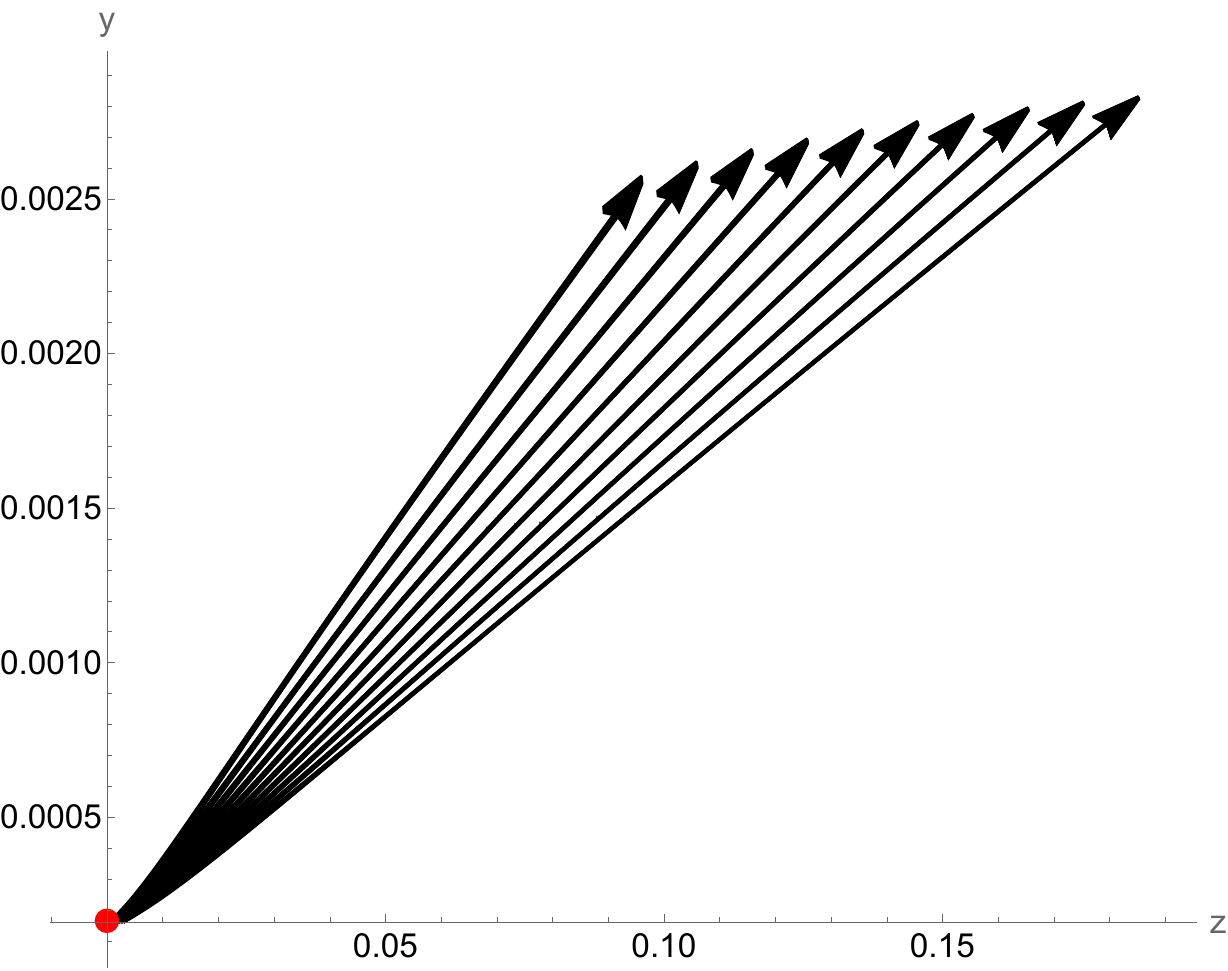} 
\caption{The phase space directions near the $P_3$ solution in the $(z, y)$ plane.}
\label{fig:p3fig2}       
\end{figure}

The last critical point found at the coordinates:

\begin{equation}
    P_3=\Big[y, z=0, \Omega_m=\frac{2 y -n y}{w} \Big], 
\end{equation}

with the existence conditions,

\begin{multline}
    w+1\neq 0\land m=\frac{n-2}{2}\land (m+1) w\neq 0
    \\
    \land 8 m^2 y+16 m n w y+48 m w y-3 n^3 w^3 y-3 n^3 w^2 y+6 n^2 w^3 y+6 n^2 w^2 y-8 n^2 w y-2 n^2 y-8 n w y+8 n y+48 w y-8 y\neq 0,
\end{multline}
describes a de-Sitter epoch ($w_{eff}=-1$) with the eigenvalues corresponding to a non-hyperbolic solution: $[0,-3,-3]$. In principle such a solution can be stable but due to the existence of one zero eigenvalue we cannot use the linear stability theory to establish the corresponding behavior. In order to obtain the physical behavior for this solution we shall rely on numerical aspects. We have displayed in Fig.~\ref{fig:p3fig2} the directions in the phase space structure near the $P_3$ solution, confirming the viability of the analytical solutions found with respect to the numerical evolutions. In this case we see that the $P_3$ point act as a repeller.

\section{The dynamical analysis for $L_m=-\rho$}
\label{ds2}
\par 
In this section we shall discuss the second scenario, where the matter Lagrangian is described by the matter density ($L_m=-\rho$), analyzing the physical effects in the phase space structure in a similar manner. Hence, the dimension--less variables are the same, 

\begin{equation}
    x=\phi,
\end{equation}

\begin{equation}
    y=\frac{V(\phi, \psi)}{6 H^2},
\end{equation}

\begin{equation}
    z=\frac{\dot{\phi}}{H},
\end{equation}

\begin{equation}
    v=\rho ~ \psi, 
\end{equation}

\begin{equation}
    \Omega_m=\frac{\rho}{3 H^2},
\end{equation}

with the following non-independent auxiliary variables:

\begin{equation}
    \Gamma=\frac{\ddot{\phi}}{H^2},
\end{equation}

\begin{equation}
    \Pi=\frac{\dot{\rho}}{H^3},
\end{equation}

\begin{equation}
    \Sigma=\frac{\dot{H}}{H^2},
\end{equation}

\begin{equation}
    E=\dot{\psi}H.
\end{equation}

The potential is described by a power law representation, $V(\phi, \psi)=V_0 \phi^n \psi^m$, with $V_0, n, m$ constant parameters. In principle, one ca extend the analysis for other potential types, changing the autonomous variables to a compatible system. In this case the Friedmann constraint equation reduces to:

\begin{equation}
    x=-\frac{1}{2} \Omega _m \left(3 v w^2+v-2\right)+y-z,
\end{equation}

with the acceleration equation, 

\begin{equation}
    \Gamma +(2 \Sigma +3) x-3 y+2 z=\frac{3}{2} \Omega _m \left(v \left(6 w^3+3 w^2+2 w+1\right)-2 w\right).
\end{equation}

These relations are complimented by the potential equations, 

\begin{equation}
    \Sigma =\frac{n y-2 x}{x},
\end{equation}

\begin{equation}
    v=\frac{2 m y}{\left(3 w^2+1\right) \Omega _m},
\end{equation}

and the auxiliary relations:

\begin{equation}
    \frac{3 m n y z}{x}+\frac{9 E (m-1) m y \Omega _m}{v}-\Pi  v \left(3 w^2+1\right)=0,
\end{equation}

\begin{equation}
    6 \Gamma +9 E \Omega _m^2-\frac{18 E m y \Omega _m}{v}+27 E w^2 \Omega _m^2+z \left(-\frac{6 n y}{x}+6 \Sigma +6\right)-2 \Pi +2 \Pi  v+6 \Pi  v w^2+12 \Sigma  x=0.
\end{equation}

For the dependent variables we get the final expressions,

\begin{equation}
    \Gamma =9 v w^3 \Omega _m+\frac{9}{2} v w^2 \Omega _m+3 v w \Omega _m+\frac{3 v \Omega _m}{2}-3 w \Omega _m+(3-2 n) y+x-2 z,
\end{equation}

\begin{equation}
    \Pi =\frac{3 m y \left(n z \left(v \left(3 w^2+1\right) \Omega _m-2 m y\right)-3 (m-1) x \left(\Omega _m \left(v \left(6 w^3+3 w^2+2 w+1\right)-2 w\right)+2 y-2 z\right)+6 (m-1) x^2\right)}{x \left(v^2 \left(3 w^2+1\right)^2 \Omega _m+2 m y \left((m-2) v \left(3 w^2+1\right)-m+1\right)\right)},
\end{equation}

\begin{equation}
    \Sigma =\frac{n y}{x}-2,
\end{equation}

\begin{equation}
    E=\frac{v \left(2 v \left(3 w^2+1\right) \left(-m n y z+3 x \left(w \Omega _m-y+z\right)+3 x^2\right)+2 m n y z-3 v^2 (2 w+1) \left(3 w^2+1\right)^2 x \Omega _m\right)}{3 x \Omega _m \left(v^2 \left(3 w^2+1\right)^2 \Omega _m+2 m y \left((m-2) v \left(3 w^2+1\right)-m+1\right)\right)}.
\end{equation}

The cosmological model is described by the following autonomous dynamical system which incorporates the physical aspects:

\begin{equation}
    \label{unu}
    \frac{dy}{dN}=\frac{y \left(-2 (m-1) y (3 m w+m-n+2)+z (-6 m w-2 m+n-4)+2 (3 m w+m+2) \Omega _m\right)}{(m-1) \left((m-1) y-\Omega _m+z\right)}, 
\end{equation}

\begin{equation}
    \frac{dz}{dN}=y \left(\frac{n z}{(m-1) y-\Omega _m+z}+6 m w+2 m-2 n+4\right)+(1-3 w) \Omega _m-z, 
\end{equation}

\begin{equation}
\label{ddoi}
    \frac{d \Omega_m}{dN}=\frac{\Omega _m \left(y (-3 m w+m+2 n+3 w-1)+(3 w-1) \Omega _m-3 w z+z\right)}{(m-1) y-\Omega _m+z},
\end{equation}

with the effective equation of state:

\begin{equation}
    w_{eff}=\frac{2 n y}{3 \left((m-1) y-\Omega _m+z\right)}+\frac{1}{3}.
\end{equation}

\par 
For the second scenario we have obtained the following critical points in the phase space structure in the case where the matter pressure is close to zero ($w \to 0$). We note that in the dust case the autonomous system is well-behaved and free of any singularities.

\par 
The first critical point $Q_1$ appears in the phase structure at the next coordinates in the case where $m=-1$, 

\begin{equation}
    Q_1=\Big[y, z=0, \Omega_m=-\frac{2 (n y+3 w y-y)}{3 w-1} \Big],
\end{equation}
with the following existence conditions:

\begin{equation}
   3 w-1\neq 0\land n w y+n y\neq 0.
\end{equation}

The physical aspects correspond to a scaling solution where the geometrical dark energy component mimic a matter epoch, $w_{eff}=w$, 

with the following eigenvalues:

\begin{equation}
    \Big[0,-\frac{\pm \sqrt{n^2 \left(24 n w-8 n+81 w^2-66 w+17\right)}-3 n w+3 n}{4 n} \Big].
\end{equation}

In the dust case ($w=0$) we have the following expressions of the eigenvalues:

\begin{equation}
    \Big[ 0,-\frac{\sqrt{(17-8 n) n^2}}{4 n}-\frac{3}{4},\frac{\sqrt{(17-8 n) n^2}}{4 n}-\frac{3}{4} \Big].
\end{equation}

\begin{figure}[h]
  \includegraphics[width=8cm]{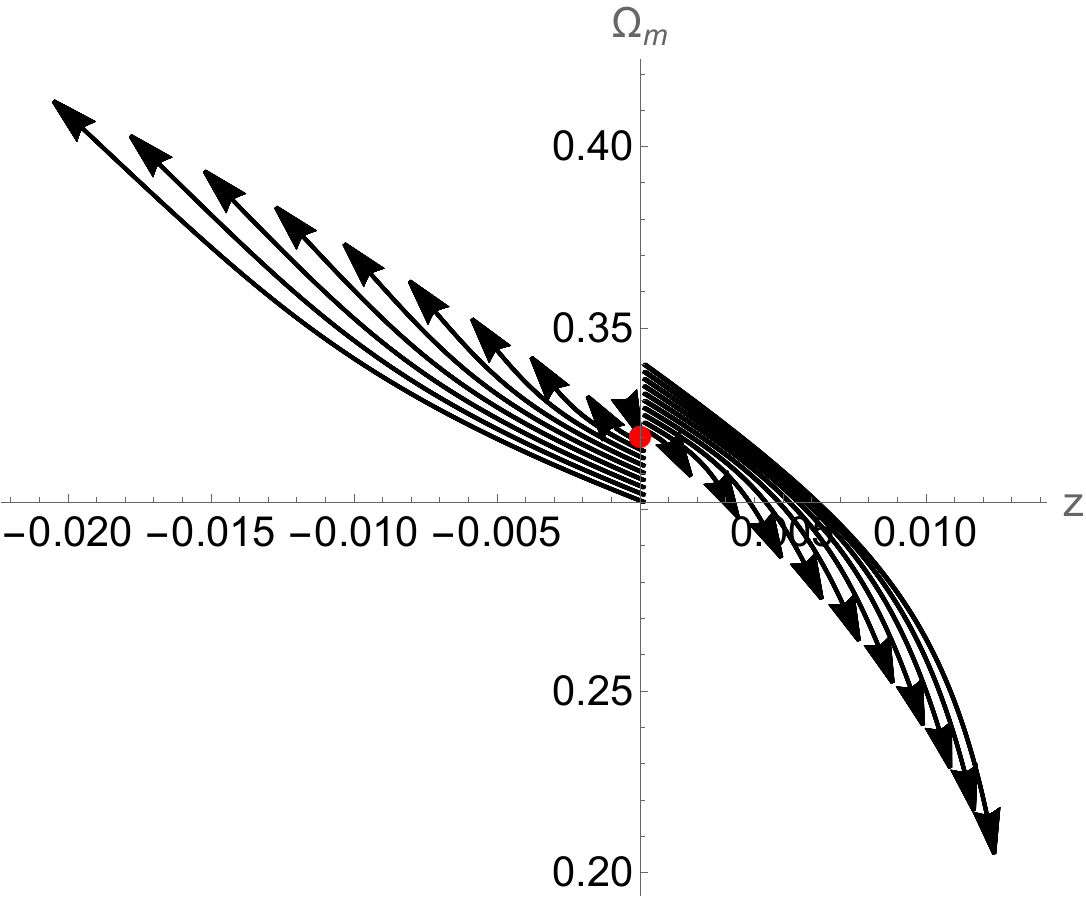} 
\caption{The phase space directions near the $Q_1$ solution in the $(z,\Omega_m)$ plane.}
\label{fig:q1fig1}       
\end{figure}

\begin{figure}[h]
  \includegraphics[width=8cm]{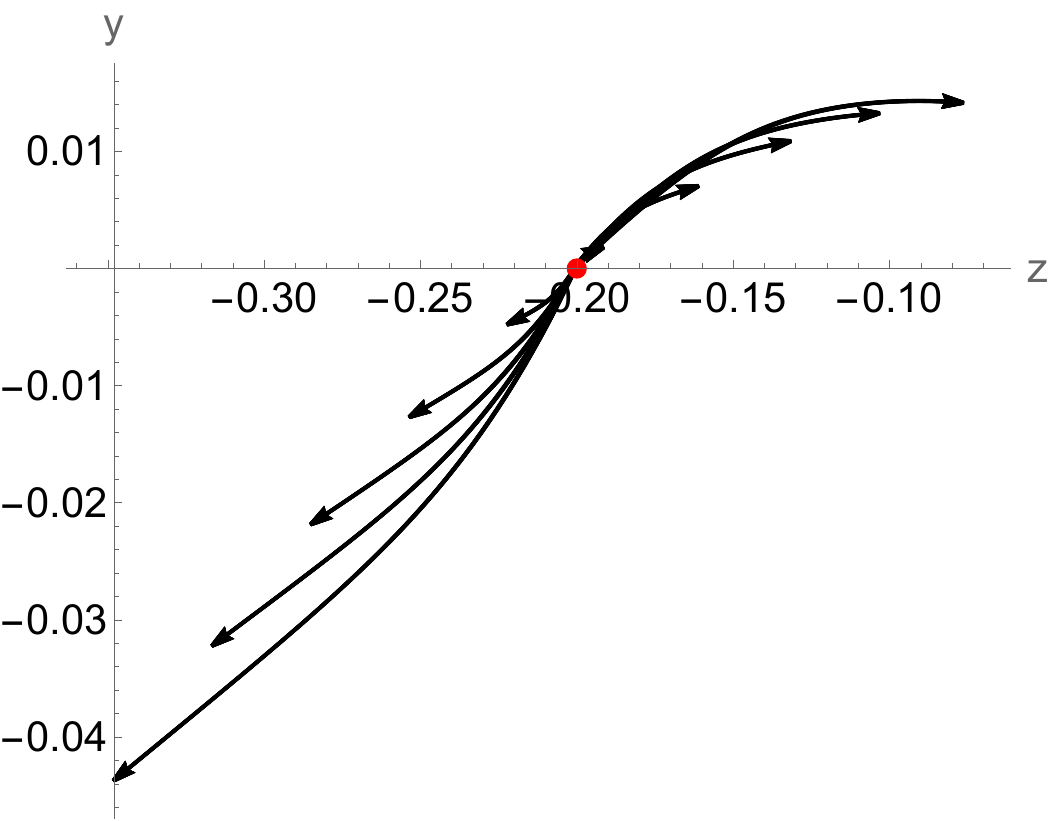} 
\caption{The phase space directions near the $Q_1$ solution in the $(z,y)$ plane.}
\label{fig:q1fig2}       
\end{figure}

\begin{figure}[h]
  \includegraphics[width=8cm]{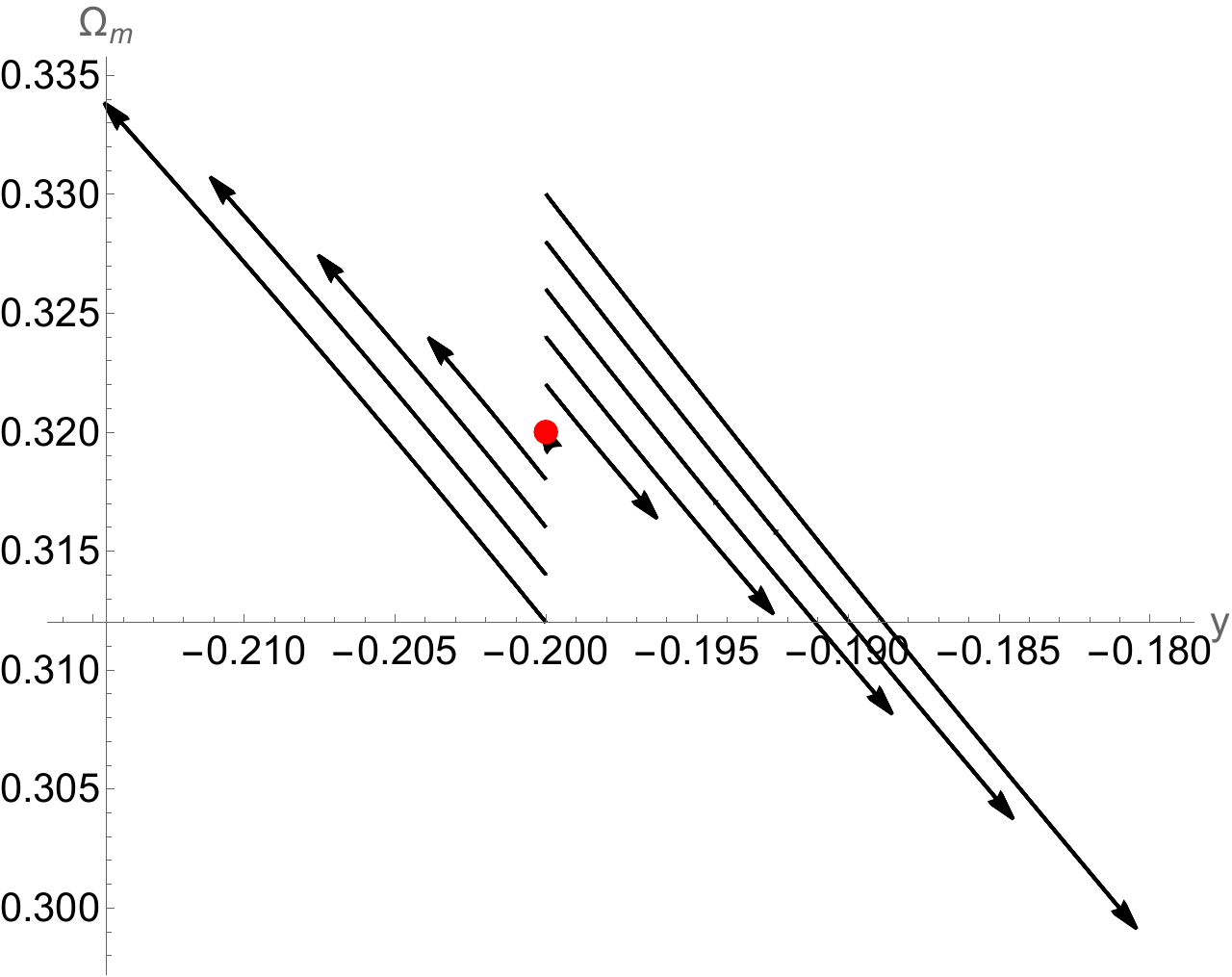} 
\caption{The phase space directions near the $Q_1$ solution in the $(y,\Omega_m)$ plane.}
\label{fig:q1fig3}       
\end{figure}

If we assume that the eigenvalues are real, we obtain the following relations which are associated to a saddle dynamical behavior, $(0<n<1\lor n<0)\land \frac{1}{2 n-2}\leq y\leq 0$, taking into account the constraint related to the matter density parameter (since $\Omega_m \in [0,1]$). The phase space directions near the $Q_1$ solutions are shown in Figs.~\ref{fig:q1fig1}--\ref{fig:q1fig3}.

\par 
The last critical point is located at the coordinates:

\begin{equation}
    Q_2=\Big[y, z=0, \Omega_m=0 \Big],
\end{equation}

found in the case where $m=\frac{n-2}{3 w+1}$ with the following existence conditions:

\begin{equation}
    3 w+1\neq 0\land n y-3 w y-3 y\neq 0.
\end{equation}

The effective equations of state, 
\begin{equation}
    w_{eff}=\frac{2 n w+n-w-1}{n-3 (w+1)},
\end{equation}
describes a possible accelerated expansion era in the dust case if the following conditions are met: $\frac{3}{2}<n<3$. The eigenvalues in the general case are the following:

\begin{equation}
    \Big[0,\frac{3 (n w+w+1)}{n-3 (w+1)},\frac{3 (w+1) (n+3 w-1)}{n-3 (w+1)} \Big].
\end{equation}

\begin{figure}[h]
  \includegraphics[width=8cm]{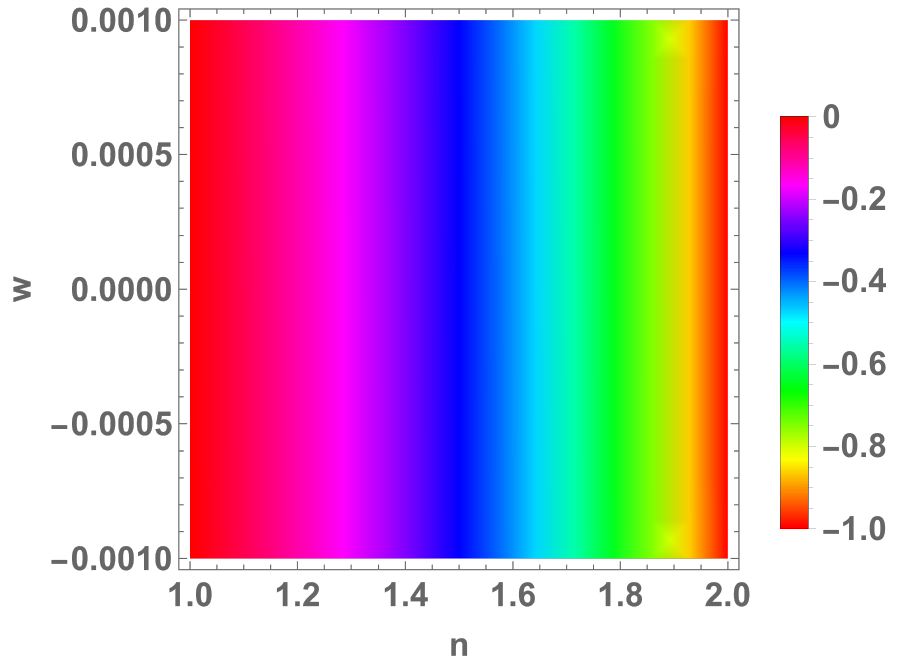} 
\caption{The effective equation of state for the $Q_2$ solution.}
\label{fig:q2weff}       
\end{figure}

We note that in the dust case a saddle behavior compatible to the accelerated expansion cannot be obtained. For this point we plot in Fig.~\ref{fig:q2weff} the value of the effective equation of state in the general case for specific intervals leading to an accelerated expansion solution.


\section{Non-exhaustive numerical analysis}
\label{nonex}
\par
\begin{figure}[h]
  \includegraphics[width=8cm]{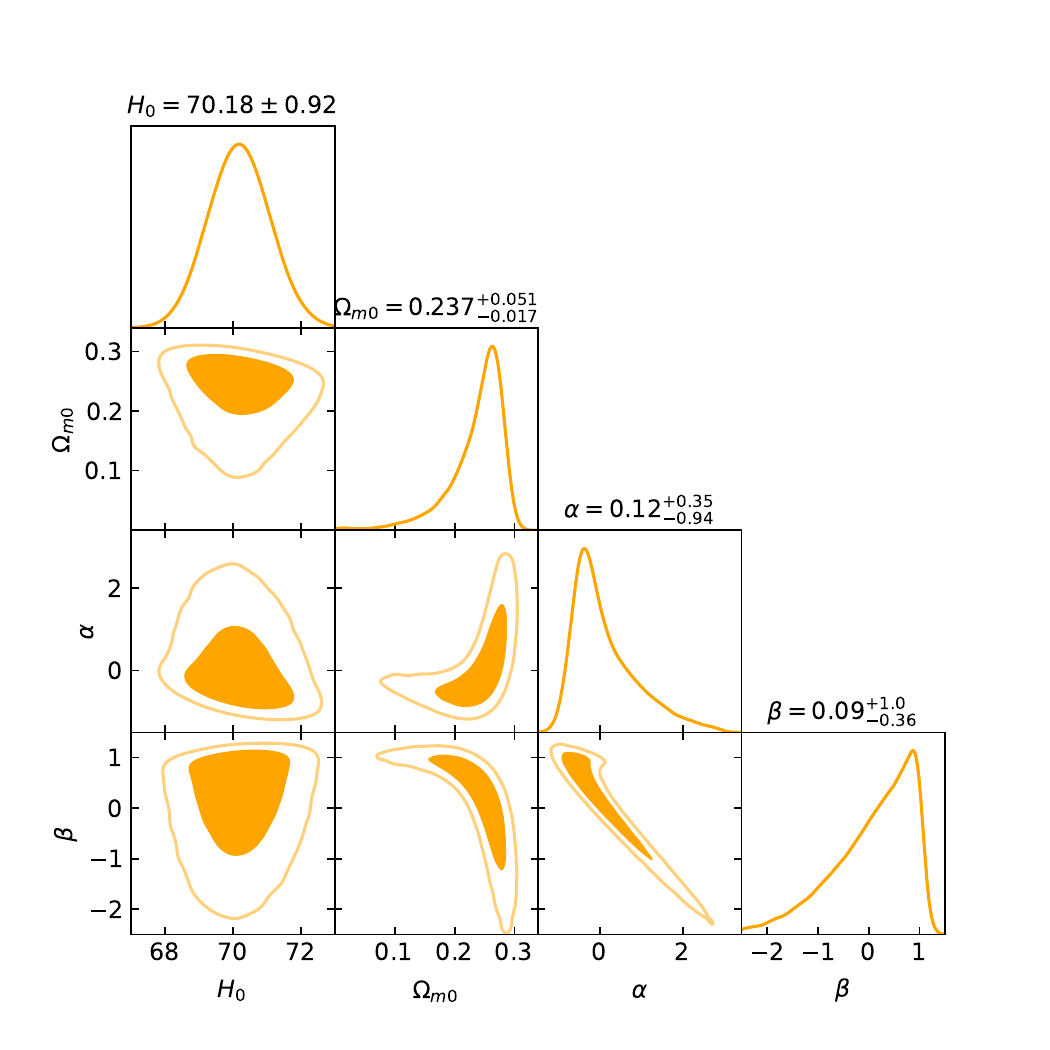} 
\caption{Posterior distributions for the $H(z)$ parameterization in the case of SnIa dataset.}
\label{fig:figsnia}       
\end{figure}

\begin{figure}[htbp]
    \centering   
    \begin{subfigure}[b]{0.34\textwidth}
        \centering
        \includegraphics[width=\textwidth]{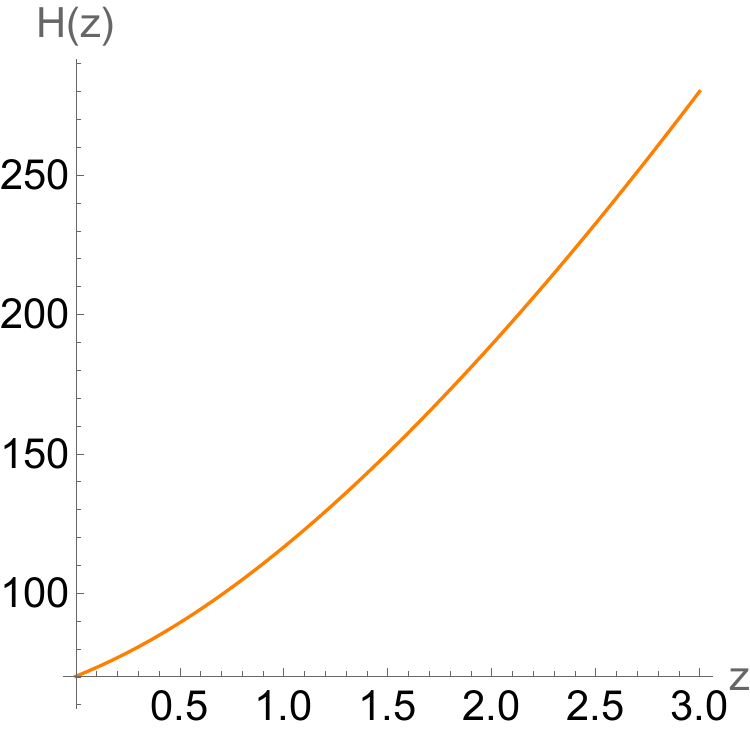}       
        \label{fig:two}
        \caption{}
    \end{subfigure}
    \hfill
    \begin{subfigure}[b]{0.34\textwidth}
        \centering
        \includegraphics[width=\textwidth]{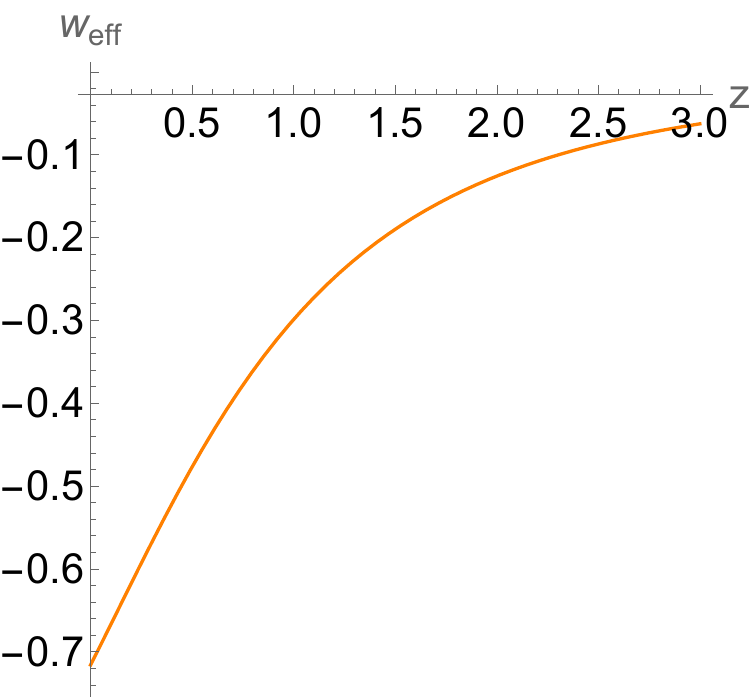}        
        \label{fig:three}
        \caption{}
    \end{subfigure}

    \caption{ a) The evolution of the Hubble rate for the best fitted parameters; b) The evolution of the effective equation of state for the best fitted parameters.}
    \label{fig:figgsnia}
\end{figure}

\begin{figure}[h]
  \includegraphics[width=8cm]{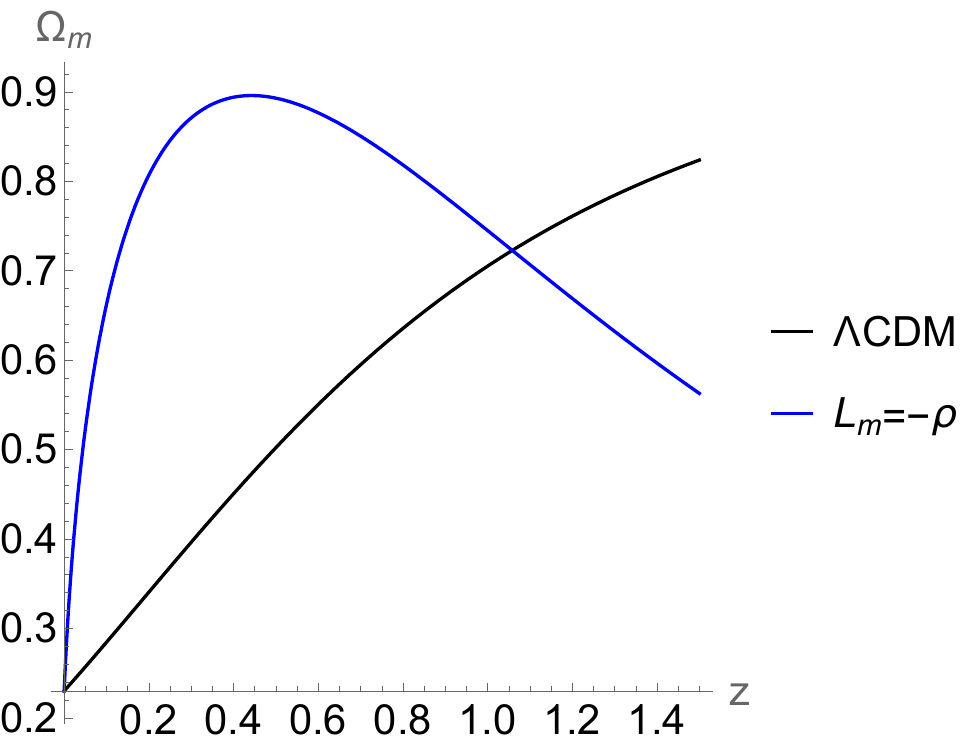} 
\caption{The evolution of the matter density parameter for the case where $L_m=-\rho$ ($V(\phi, \psi)=V_0 \phi \psi, V_0=1, w=0$).}
\label{fig:lmegalrho}       
\end{figure}

\begin{figure}[h]
  \includegraphics[width=8cm]{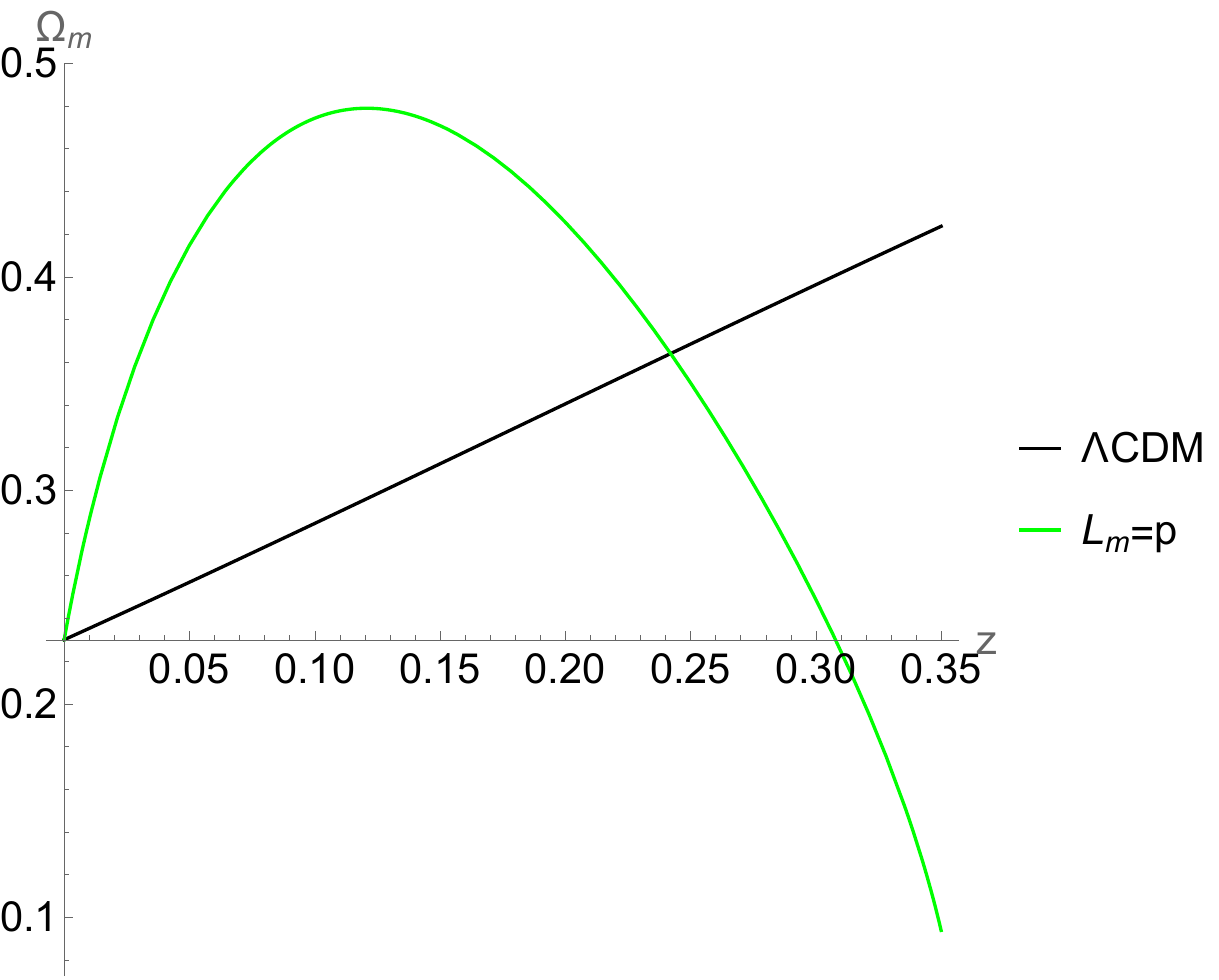} 
\caption{The evolution of the matter density parameter in the case where $L_m=p$ ($V(\phi, \psi)=V_0 \phi \psi, V_0=1, w=0.1$).}
\label{fig:lmegalp}       
\end{figure}

 
In this section we shall discuss possible physical effects, by taking into account a numerical approach. In order to analyze  physical effects, we shall first investigate a specific parameterization for the Hubble parameter. Hence, we shall consider the following parameterization \cite{Mamon2017}:

\begin{equation}
    H(z)=H_0\sqrt{\Omega_{m0}(1+z)^3+(1-\Omega_{m0})(1+z)^\alpha e^{\beta z}},
\end{equation}

constraining the model's parameters $\{ H_0, \Omega_{m0}, \alpha, \beta\}$. The Hubble rate is evaluated by including the supernovae of type Ia in the definition of $\chi^2$, encapsulating cosmic chronometers, baryon acoustic oscillations, and supernovae datasets. For the numerical procedure, see Ref.~\cite{Marciu:2023hdb} and references therein. The results of the MCMC procedure are depicted in Fig.~\ref{fig:figsnia}, with the best fitted values of specific parameters displayed. For the best fitted values we have plotted the Hubble expansion rate and the effective (total) equation of state in Fig.~\ref{fig:figgsnia}. In this case, we can note that the evolution corresponds to a quintessence behavior at late times. Next, we have used this numerical ansatz for the Hubble expansion rate in the corresponding relations, the Friedmann expressions and the potential equations. The results of the numerical evaluations are described in Figs.~\ref{fig:lmegalrho}--\ref{fig:lmegalp} for specific initial conditions. In the case of $L_m=-\rho$ we can see that the matter density parameter evolves differently with respect to the $\Lambda$CDM model, the matter domination era is attained at late times, for small redshifts. Moreover, after reaching a matter domination era in the recent past, the evolution describes a Universe dominated by a geometrical dark energy component in the early times, for high redshift. In the second scenario, where $L_m=p$ the evolution has a similar behavior, noting that a higher discrepancy with respect to the $\Lambda$CDM model is observed. The matter density parameter evolves towards an increased state stage where the matter density parameter is higher than today. However, after reaching a maximum stage, the matter density parameter drops and the Universe is dominated by a geometrical dark energy component. Note that in order to obtain the previous mentioned evolutions we have fine--tuned the model's parameters and the initial values for the final system of differential equations. In the case where $L_m=p$, the dark matter component is a warm constituent \cite{Yao:2025kuz}, having a nonzero pressure, since the cold dark matter scenario leads to a singularity problem. Finally, we note that the evolution of the matter density parameters discussed earlier confirm the energy flow between the matter component and the dark energy element, an exchange of geometrical nature.

\section{Conclusions}
\label{conclusions}
\par 
In this paper we have revisited the energy-momentum squared gravity, a cosmological theory which extends the fundamental Einstein-Hilbert action by considering a specific term based on the energy-momentum tensor. In this context we have taken into account in the derivation of the field equations the double variation of the matter Lagrangian with respect to the metric, a component which has been neglected in the previous studies. Hence, in the first part of the manuscript we have discussed the specific features of this term and the physical implications which are based on thermodynamical grounds. Afterwards, we have obtained the scalar tensor representation, by introducing two additional fields in the action which can count for the dynamical representation of the previous mentioned theory. 
\par 
Furthermore, by taking into account two Lagrangian densities of the matter (dark matter) component, we have analyzed the physical implications by relying on the linear stability theory. We have introduced specific dimension-less auxiliary variables in each case, analyzing the physical features of the phase space structure. In these cases, we have discussed the emergence of several cosmological epochs, which include matter domination, and the late time era close to a de--Sitter behavior.  
\par 
In the literature various Lagrangian densities for the matter component have been proposed, the pressure ($L_m=p$), the specific density ($L_m=-\rho$), and the trace of the energy-momentum tensor ($L_m=T$) \cite{Minazzoli:2025gyw}. In this regard, we have considered the two most common cases, when the Lagrangian density is defined by the pressure, and a second case where the density describes the Lagrangian density. For each of these cases we have analyzed the phase space structure and the emergence of the critical points, discussing the relevant dynamical aspects. These aspects further constrained the energy-momentum squared gravity theory, adding physical features which are connected to the dynamical analysis. 

\par 
The present paper can be extended in various ways. For example, a full dynamical system analysis of the complete theory which includes the second derivative of the matter Lagrangian with respect to the inverse metric can be considered. Another aspect would be related to the compatibility with cosmological observations, by taking into account the validity of the latter theory with respect to the cosmic chronometers, baryonic acoustic oscillation, supernovae, and quasi-stellar objects. All of these can add further constraints of the model, validating the  complete theory. Furthermore, different aspects related to the neutron star structure can be considered, by applying the same formalism in the context of these astrophysical objects. As can be noted, the present manuscript can be extended in several directions which are open for investigation. Some of these aspects are left for future projects and shall be presented in the future.

\bibliography{sorsamp}

\end{document}